\documentclass[prb, twocolumn]{revtex4-1}
\usepackage{amsmath}
\usepackage{amssymb}
\usepackage{amsfonts}
\usepackage[pdftex]{graphicx}
\usepackage[utf8]{inputenc}

\begin{document}

\title{Statistical distribution of thermal vacancies close to the melting point}

\author{Mar\'{\i}a Jos\'e Pozo}
\email{mariaj.pozo@gmail.com}
\affiliation{Grupo de Nanomateriales, Departamento de F\'{i}sica, Facultad de 
Ciencias, Universidad de Chile, Casilla 653, Santiago, Chile}
\author{Sergio Davis}
\homepage{http://www.gnm.cl/~sdavis}
\email{sdavis@gnm.cl}
\affiliation{Grupo de Nanomateriales, Departamento de F\'{i}sica, Facultad de 
Ciencias, Universidad de Chile, Casilla 653, Santiago, Chile}
\author{Joaqu\'{\i}n Peralta}
\affiliation{Departamento de Ciencias F\'isicas, Facultad de Ciencias Exactas,  
Universidad Andr\'es Bello, Santiago, Chile}
\email{joaquin.peralta@unab.cl}

\date{\today}

\begin{abstract}
A detailed description of the statistical distribution of thermal vacancies near the melting 
point is presented, using copper as an example. 
As the temperature is increased, the average number of thermal vacancies generated by
atoms migrating to neighboring sites also increase, according to Arrhenius'
law. We present for the first time a model for the distribution of thermal
vacancies, which according to our results follow a Gamma distribution.
All the simulations are carried out by classical molecular dynamics and the 
recognition of vacancies is achieved via a recently developed algorithm. Our
results could be useful in the further development of a theory explaining the
mechanism of homogeneous melting, which seems to be mediated (at least in part)
by the accumulation of thermal vacancies near the melting point.
\end{abstract}



\maketitle

\section{Introduction}

Understanding the production of thermal vacancies due to atomic migration near
the melting temperature $T_m$ should provide relevant information on the melting process 
itself. Recent studies involving computer simulation~\cite{Delogu2005, Bai2008, Gallington2010, Davis2011,
Zhang2013} have connected the catastrophic collapse of the crystal in
homogeneous melting to a collective (\emph{ring}-like) movement of atoms due to
thermally produced vacancies. In order to construct a quantitative model, however, there is a 
key piece missing: the statistical distribution of thermal vacancies at a given
temperature $T$.  

For temperatures close to $T_m$, the expected concentration of vacancies is between
10$^{-3}$ and 10$^{-4}$ for metals~\cite{Hillert2007}. However, from this point on it is naturally 
expected (and indeed true) that the number of thermal vacancies increases. Another common assumption is that, at a
fixed temperature $T$, the concentration of vacancies $f_v=n_v/N$ (where $n_v$
is the number of vacancies and $N$ the total number of atoms) is \emph{normally distributed}
around an average value $\big<f_v\big>$ which follows Arrhenius' law, 

\begin{equation}
\big<f_v\big>_T = \exp(-E_v/k_B T)
\label{eq_arr}
\end{equation}
where $E_v$ is the free energy of formation of thermal vacancies~\cite{Hillert2007}.

In this work we provide evidence from atomistic computer simulations supporting
a Gamma model for the concentration of thermal vacancies in copper near $T_m$ (but below $T_{LS}$).

The work is organized as follows. Section~\ref{Simulation} shows a detailed 
description of the molecular dynamics and vacancy recognition procedures. Section~\ref{Distribution} 
shows the details of the inference process employed for the statistical
comparison of the Gamma and normal models for vacancy distribution.
Section~\ref{Concluding} comments on the scope and implications of our results.

\section{Simulation techniques}
\label{Simulation}

\begin{figure}[!ht]
\includegraphics[scale=.16]{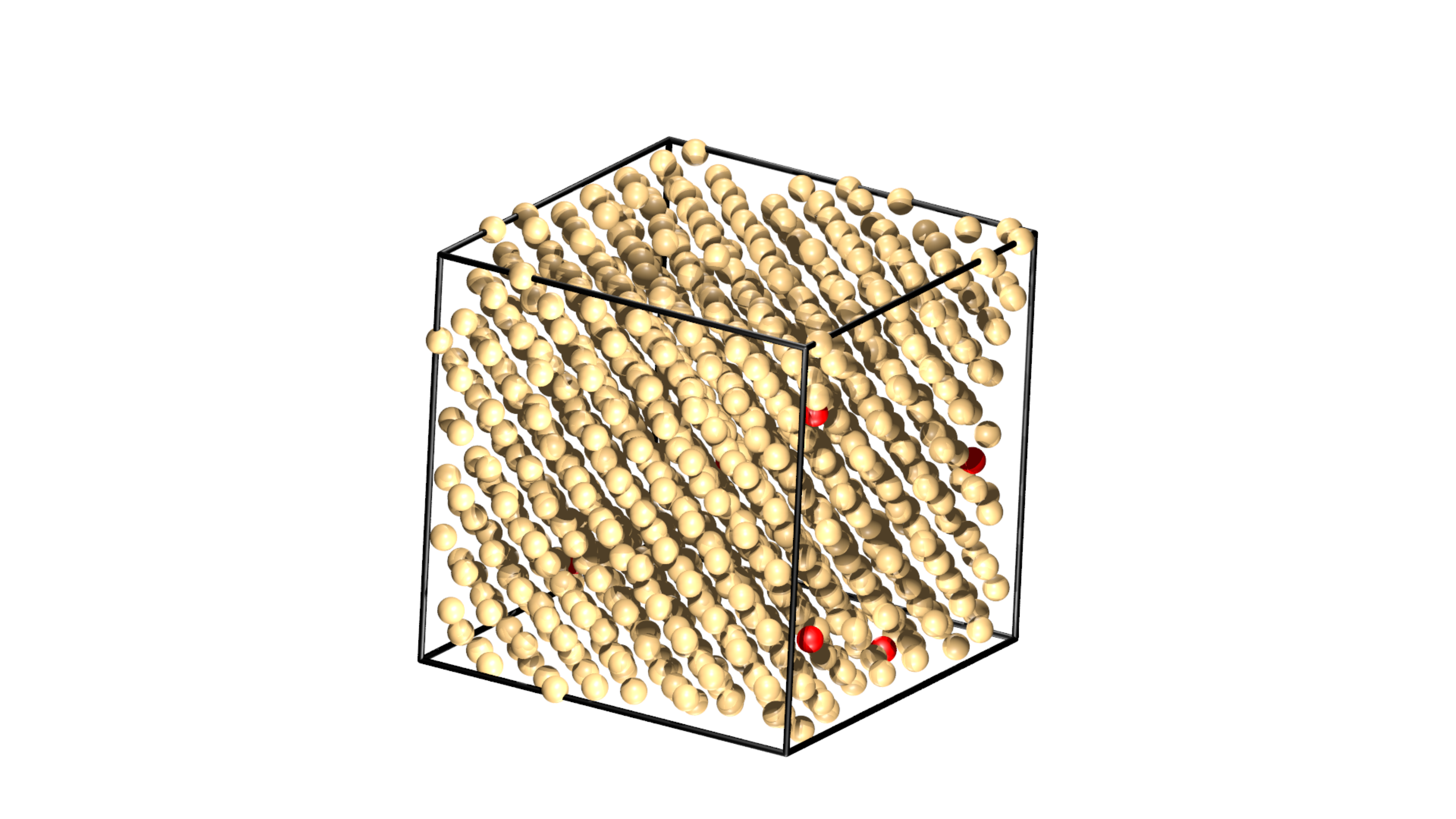}
\vspace{15pt}
\caption{Copper structure snapshot from the MD simulation at $T=$1400 K. The structure has 1372 
atoms, with an initial FCC crystalline structure. The red spheres represent
thermal vacancies located with the \emph{Search-and-Fill} algorithm.}
\label{Cu-fig}
\end{figure}

The simulations were performed using an FCC copper structure composed of 1372
atoms (7x7x7 unit cells), using a lattice parameter of $a=3.61$\AA ~in a cubic cell of length $L=25.27$\AA, as 
shown in Fig. ~\ref{Cu-fig}. The lattice parameter corresponds to room
pressure copper. 

Classical molecular dynamics (MD) simulations were performed in the microcanonical ensemble 
(i.e., with fixed $N$, $V$ and $E$) using the LPMD software package~\cite{Davis2010}. 
We described the interatomic interactions in copper by the Sutton-Chen potential with the usual
parameterization~\cite{Sutton1990}. 

Simulations at four different initial temperatures, 2300 K, 2500 K, 2600 K, and 2700 K, were performed 
for 50 ps each, with a timestep $\Delta$t$=1$ fs. In all cases the initial
temperature $T_0$ was set to be about twice the target temperature $T$, and the ideal crystalline
structure is used as the initial positions. In this manner equilibration is achieved without the use of
thermostats which could distort the natural dynamics of the system. This is the
same microcanonical approach used in the Z-method~\cite{Belonoshko2006}. The
temperatures were chosen close to the experimental melting point of copper,
$T_m \sim $1360 K, in order to have a non-zero probability of observing thermal
vacancies given the size of the system.

\begin{figure}[!ht]
\includegraphics[scale=.5]{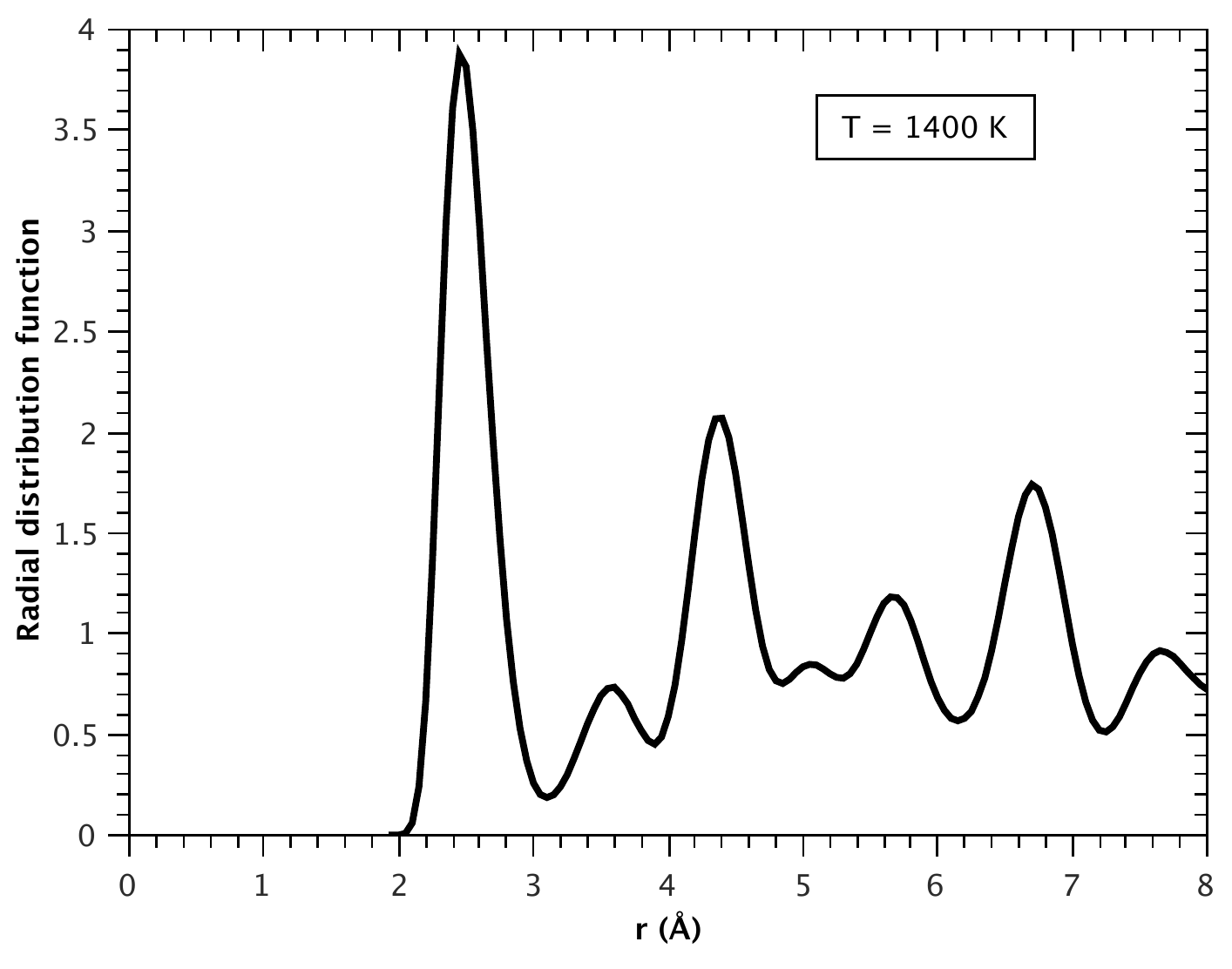}
\caption{Pair distribution function for copper at $T=$1400 K. It shows the
structure is still an FCC solid, with the usual broadening of the peaks due to
temperature.}
\label{gdr}
\end{figure}

We computed the radial distribution function $g(r)$ for all the temperatures in
order to check that we indeed have a solid structure in all cases. Figure
\ref{gdr} shows the $g(r)$ for the case of $T=$1400 K, all the other
temperatures being almost identical. In this figure we can see that the
nearest-neighbors peak is located around $r=$2.5 \AA~, which gives the
approximate radius of a spherical vacancy to be close to 1.25 \AA.

To determine the number of vacancies generated during the simulations, the
\emph{Search-and-Fill} algorithm~\cite{Davis2011b} was used. This technique
generates virtual spheres in the simulation cell (of radius $R_0$) and tries to
place them with minimum overlap with the atoms. Every site where a virtual
sphere can fit with an overlap below a threshold $\Omega$ is identified as a vacancy 
and the site is filled (i.e., the site is not considered empty for the purposes
of locating the next vacancy). In the particular case of copper we used the values of 
$R_0=$1.275 \AA ~and the threshold overlap parameter $\Omega=$0.4.

The average results obtained from the vacancy recognition procedure are presented in 
Table~\ref{tab:vac}. As expected, the concentration of vacancies increases with temperature.

\begin{table}
\begin{tabular}{cc}\\\hline
 T (K) & $\big<f_v\big>$ (10$^{-3}$) \\ \hline 
 1200 & 2.06195 \\
 1300 & 3.22012 \\
 1350 & 3.97303 \\
 1400 & 4.62318 \\ \hline
\end{tabular}
\caption{Values of the average concentration of thermal vacancies for several
temperatures.}
\label{tab:vac}
\end{table}

\section{Vacancy distribution}
\label{Distribution}

\begin{figure}
\includegraphics[scale=.6]{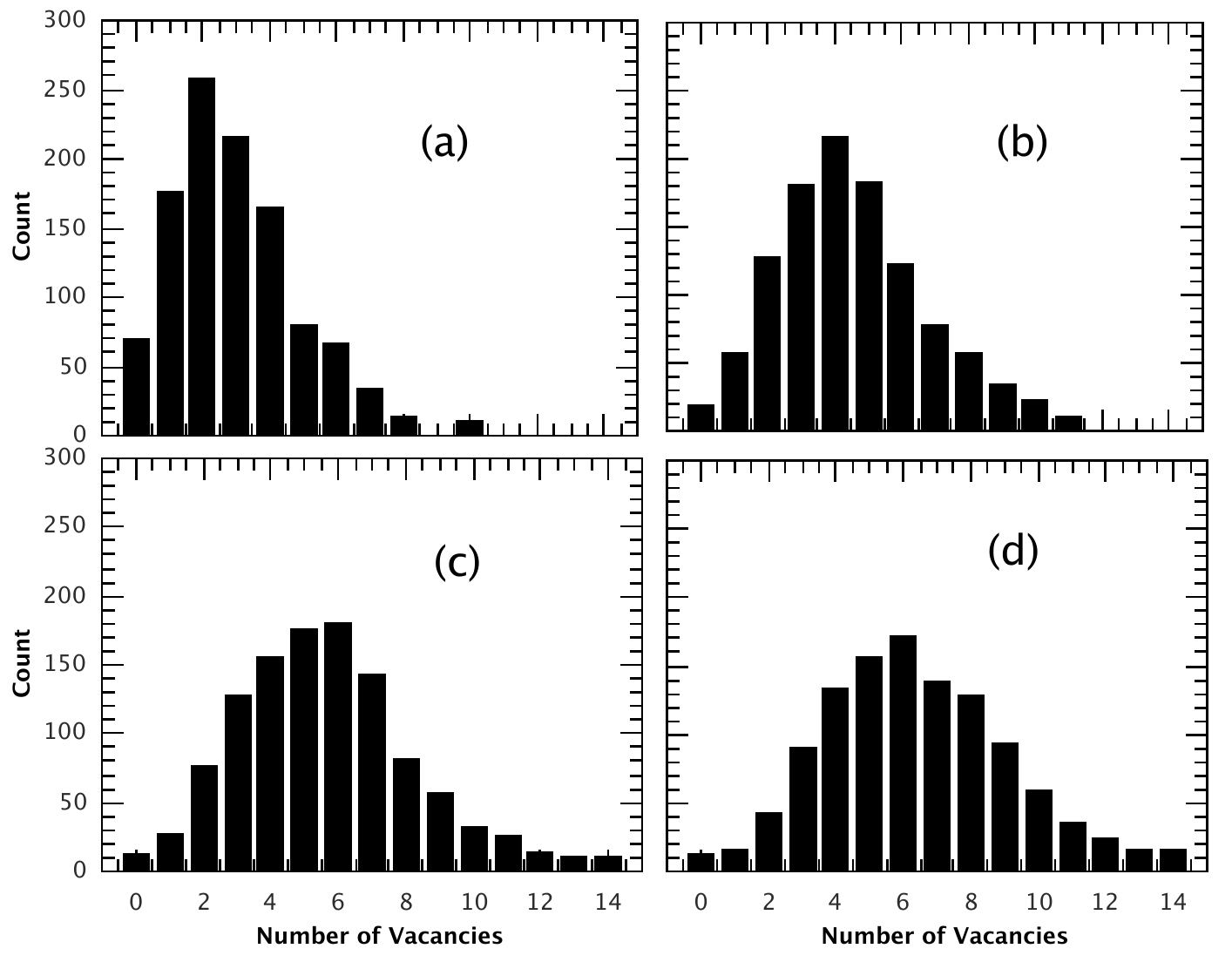}
\caption{Vacancy distribution corresponding to each temperature. The histogram
represents the frequency of each number of vacancies present in a sample of 1372
atoms.}
\label{histograms}
\end{figure}

To determine how the vacancies are distributed in the sample we evaluate the 
number of vacancies in each time step during the simulation. The results of this
vacancy count are organized in a histogram. These histograms for four different 
temperatures are displayed in Fig. \ref{histograms}.

We propose two models for the probability distribution of vacancies, a Gaussian
distribution,

\begin{equation}
P(f_v|\mu, \sigma) = \frac{1}{\sqrt{2\pi}\sigma}\exp(-\frac{1}{2\sigma^2}(f_v-\mu)^2)
\end{equation}
where $\mu$ and $\sigma^2$ are the mean and variance of $f_v$, respectively, and
a Gamma distribution, given by

\begin{equation}
P(f_v|k, \theta) = \frac{1}{\Gamma(k)\theta^k}f_v^{k-1}\exp(-f_v/\theta)
\end{equation}
where $k$ and $\theta$ are its shape and scale parameters, respectively. In both cases 
the parameters of the distributions are functions of $T$. In order to compare both models 
using our simulated data for each temperature, the Bayesian Information Criterion (BIC)~\cite{Schwarz1978}, defined as

\begin{equation}
\text{BIC} = -2\ln L(\mathbf{\hat{\lambda}_0)}+n_p\ln N,
\end{equation}
has been used, where $L(\mathbf{\lambda})$ is the likelihood function for the model,
$\mathbf{\lambda_0}$ are the most probable parameters according to the maximum likelihood
method, $n_p$ is the number of parameters in the model and $N$ is the number of
data points. In this method, the lower the value of BIC, the better (the model
gives a better fit to the data). In our case, the second term is the same for
both models, and so the comparison reduces to a maximum likelihood ratio. 
The results for each temperature are displayed in Table~\ref{tab:bic}.

\begin{table}
\begin{tabular}{ccc}\\\hline
 T (K) & BIC for Gamma & BIC for normal \\ \hline 
 1200 & 3852.5 & 3936.1 \\
 1300 & 4273.7 & 4302.6 \\
 1350 & 4475.4 & 4486.1 \\
 1400 & 4695.9 & 4725.2 \\ \hline
\end{tabular}
\caption{Values of the Bayesian Information Criterion (BIC) for the Gamma and
normal models applied to the concentration of thermal vacancies for several
temperatures. In all cases the Gamma model is selected (has lower BIC) over the
normal.}
\label{tab:bic}
\end{table}

For all temperatures, the Gamma model is to be preferred over the normal model,
and in all cases the difference in BIC is larger than 10 points, giving for each
temperature odds in favor of the Gamma model higher than 140:1. We therefore
conclude that the evidence in favor of the Gamma model is statistically
conclusive for our data (see for instance Raftery~\cite{Raftery1995} for the
statistical significance of Bayes factors and BIC differences).

\begin{figure}
\begin{center}
\includegraphics[scale=0.15]{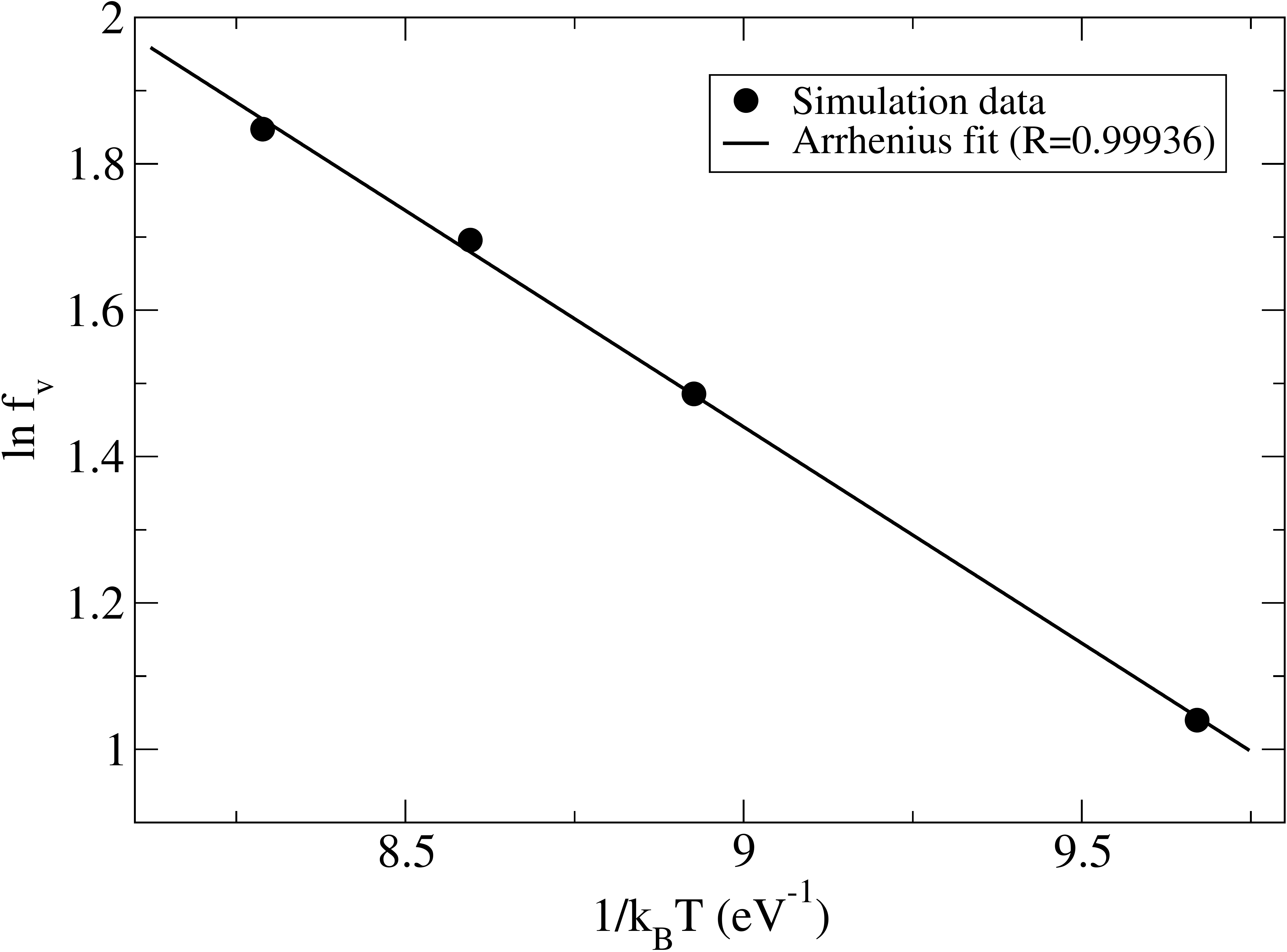}
\end{center}
\caption{Arrhenius law for the average concentration of vacancies.}
\label{arrhenius}
\end{figure}

The Arrhenius law (Eq. \ref{eq_arr}) for our calculated averages is displayed in
Fig. \ref{arrhenius}, which corresponds to an activation energy $E_v \sim
0.59112$ eV, lower than previous experimental results on intrinsic
vacancies~\cite{Simmons1963, Triftshauser1975} (reporting values around 1.0 eV).

This is to be expected, as the formation of an intrinsic vacancy is more costly,
due to it involving the removal of an atom from the surface, while the thermal
vacancy is actually a vacancy-interstitial pair, and involves just a local rearrangement 
of atomic overlap. 

With the obtained value of $E_v$, the concentration of vacancies at the
experimental melting point is $f_v(T_m) \sim $ 4$\times$10$^{-3}$, in agreement
with known values for metals~\cite{Hillert2007}.

\section{Concluding remarks}
\label{Concluding}

The formation of thermal vacancies is a stochastic phenomenon which, however,
seems to follow a well defined statistical distribution. We provide evidence supporting a 
Gamma distribution with long tails instead of the more common normal distribution, which
increases the probability of larger concentrations at a given mean (extreme
events). This has implications for the modelling of the homogeneous melting
process which is dependent on the formation of thermal vacancies and their
mobility, as the normal model would underestimate the probability of a critical
vacancy concentration. The procedure employed in this work is capable of
determining the free energy of formation of vacancies by performing molecular
dynamics simulations or even Monte Carlo simulations.

\section{Acknowledgements}

SD \& JP gratefully acknowledge funding from FONDECYT grant 1140514.

\bibliography{vacdist}
\bibliographystyle{apsrev}

\end{document}